\documentclass[aps,prl,twocolumn,showpacs,superscriptaddress,groupedaddress]{revtex4}  
\usepackage{graphicx}  
\usepackage{dcolumn}   
\usepackage{bm}        
\usepackage{amssymb}   
\usepackage[a4paper, top=2cm, bottom=2cm, left=1.9cm, right=1.9cm]{geometry}
\begin{document}


\title{Metallic Icosahedron Phase of Sodium at Terapascal Pressures}

\author{Yinwei Li} \affiliation{School of Physics and Electronic Engineering, Jiangsu Normal University, Xuzhou 221116, China} \affiliation{State Key Laboratory of Superhard Materials, Jilin University, Changchun 130012, China}
\author{Yanchao Wang} \affiliation{State Key Laboratory of Superhard Materials, Jilin University, Changchun 130012, China}
\author{Chris J. Pickard} \affiliation{Department of Physics and Astronomy, University College London, Gower Street, London WC1E 6BT, United Kingdom}
\author{Richard J. Needs} \affiliation{Theory of Condensed Matter Group, Cavendish Laboratory, J J Thomson Avenue, Cambridge CB3 0HE, United Kingdom}
\author{Yi Wang} \affiliation{Natural Science Research Center, Academy of Fundamental and Interdisciplinary Sciences, Harbin Institute of Technology, Harbin 150080, China}
\author{Yanming Ma}\email{mym@jlu.edu.cn} \affiliation{State Key Laboratory of Superhard Materials, Jilin University, Changchun 130012, China}

\begin{abstract}
Alkali metals exhibit unexpected structures and electronic behavior at high pressures. Compression of metallic sodium (Na) to 200 GPa leads to the stability of a wide-band-gap insulator with the double hexagonal hP4 structure. Post-hP4 structures remain unexplored, but they are important for addressing the question of the pressure at which Na reverts to a metal. Here we report the reentrant metallicity of Na at the very high pressure of 15.5 terapascal (TPa), predicted using first-principles structure searching simulations. Na is therefore insulating over the large pressure range of 0.2-15.5 TPa. Unusually, Na adopts an oP8 structure at pressures of 117-125 GPa, and the same oP8 structure at 1.75-15.5 TPa. Metallization of Na occurs on formation of a stable and striking body-centered cubic cI24 electride structure consisting of Na$_{12}$ icosahedra, each housing at its center about one electron which is not associated with any Na ions.
\end{abstract}

\maketitle

Alkali metals have long been known to possess simple electronic structures at ambient pressure that are well explained by a nearly-free-electron model. However, the simple low-pressure structures of alkali metals do not remain simple upon compression. Rich and complex phases and remarkable physical phenomena have been observed, such as greatly increased electrical resistivity\cite{1}, enhanced superconductivity\cite{2,3}, unusually low melting temperatures\cite{4,5} and metal-insulator/semiconductor transitions\cite{6,7}.

The metal-insulator transitions in Na and lithium  (Li)\cite{6,7} are among the most fascinating observations at high pressure. Neaton and Ashcroft predicted that upon compression Li\cite{8} and Na\cite{9} might transform into atomically-paired structures. If a paired structure was to be adopted, both Na and Li would have the potential to be semiconducting. These predictions have stimulated numerous studies of the high-pressure structures of Li and Na (e.g., Refs. \cite{6,7,10,11,12,13,14}).

The metal-insulator transition in Na was established by a joint theoretical and experimental effort\cite{7} and confirmed by further experiments\cite{15}. Na is predicted and observed to transform into an optically transparent phase at above 200 GPa\cite{7}. This phase was predicted and experimentally confirmed to have a double hexagonal hP4 structure\cite{7}, with a remarkably large bandgap reaching 6.5 eV at 600 GPa. A metal-semiconductor transition was observed in Li at 80 GPa. The semiconducting structure of dense Li remained unsolved despite considerable attention (e.g., Refs.\cite{11,12,13,14}). Blind crystal structure prediction calculations[12] on dense Li predicted a complex semiconducting oC40 structure. Independent experimental and theoretical studies also found that semiconducting Li adopts the oC40 structure\cite{16}.

Analysis of the charge density\cite{7,11,12} shows that the emergence of an insulating state in dense Na/Li is, however, not attributable to atomic pairing, but rather to strong localization of valence electrons within lattice voids. Notably, the insulating phases of Na and Li can be regarded as prototypical examples of electrides\cite{17}, in which the ionic cores play the role of cations, and interstitial electrons form the anions\cite{7,11,12}. Stable electride structures have also been predicted in other elements at high pressures, e.g., Al\cite{18}, Mg\cite{19}, Ca\cite{20} and K\cite{21}. The formation of an electride in Na at high pressures is favorable because it reduces the kinetic energy of the higher-energy electrons. These electronic orbitals are forced to oscillate rapidly close to the atomic cores due to Pauli repulsion, but if they move away from the cores the oscillations and hence the kinetic energy can be reduced\cite{17,18}. The electrides have open structures if one considers only the ions, but when the interstitial electrons are considered as anions the structures make good chemical sense as ionic solids.

Pressure-induced metal-insulator transitions have been predicted in other materials at TPa pressures. For example, it was suggested that Mg, a good metal at normal pressures, would transform from a metal to a semimetal at 24 TPa\cite{Mg}, and Ni was predicted to transform to an insulator at 34 TPa\cite{Ni}. The metal-insulator transition in Ni is driven by the complete 4s$\rightarrow$3d charge transfer under pressure, which is different from the electron localization found in Li and Na.

Metallization is presumed to be the ultimate fate of all materials under sufficiently strong compression\cite{22}. The insulating phase of Ni has been predicted to revert to metallic behavior at 51 TPa\cite{Ni}. Recent electrical resistance measurements have revealed that semiconducting Li reverts to a "poor metal" above 120 GPa\cite{10}. The question of the pressure at which Na will also revert to a metal remains unanswered. Determining the ground state structures of Na at high pressures provides the key to whether insulating or metallic behavior is observed. At room temperature, Na undergoes pressure-induced phase transitions with an established sequence of bcc $\rightarrow$ fcc (65 GPa)\cite{23} $\rightarrow$ cI16\ (103 GPa)\cite{24,25,26} $\rightarrow$ oP8 (117 GPa)\cite{24} $\rightarrow$ tI19 (125 GPa)\cite{24} $\rightarrow$ hP4 (200 GPa)\cite{7}. The hP4 structure is the highest-pressure phase of Na known so far.

In this letter, we report the reentrant metallicity of Na at a surprisingly high pressure of 15.5 TPa, predicted via extensive structure searching in conjunction with first-principles calculations. We predict that Na will show insulating behavior over a remarkably large pressure range of 0.2-15.5 TPa and will exhibit a phase transition to a second insulating orthorhombic oP8 structure at 1.75 TPa. Metallization of Na appears with the formation of a striking body-centered cubic (bcc) cI24 structure whose lattice sites are populated by Na$_{12}$ icosahedra, each containing about one electron which is not associated with any Na ions.

Our structure searching calculations with simulation cells containing up to 24 atoms were performed in a wide pressure range (0.4-20 TPa). We used the efficient CALYPSO\cite{27,28} and AIRSS\cite{29,30} methods, which have both been successfully applied to investigating structures of materials at high pressures\cite{12,29,31,32,33,34,35,36,37,38,39,40}. TPa pressures can now be achieved with dynamic ramped compression. In particular, diamond has been studied experimentally up to 5 TPa\cite{41}. The density functional theory\cite{42} (DFT) calculations were performed using the CASTEP plane-wave code\cite{43} with the Perdew-Burke-Ernzerhof\cite{44} (PBE) generalized gradient approximation (GGA) functional. We performed calculations with both norm-conserving and ultrasoft Na pseudopotentials, in which only the 1\emph{s} energy level was treated as core. An ultrasoft pseudopotential was used for the structure searches, with a plane-wave cutoff energy of 910 eV and a Brillouin-zone integration grid spacing of 2$\pi \times 0.05{\AA}^{-1}$. The structures obtained were re-optimized at a higher level of accuracy for both the norm-conserving and ultrasoft pseudopotentials. Here, we present only norm-conserving results since the two pseudopotentials led to consistent conclusions. A cutoff energy of 1633 eV and a k-point grid spacing of 2$\pi \times 0.03{\AA}^{-1}$ were used in the calculations with norm-conserving pseudopotentials. The phonon dispersion curves were computed by a finite displacement method as implemented in the CASTEP code.

\begin{figure}
  \begin{center}
  \includegraphics[width=1.0\columnwidth]{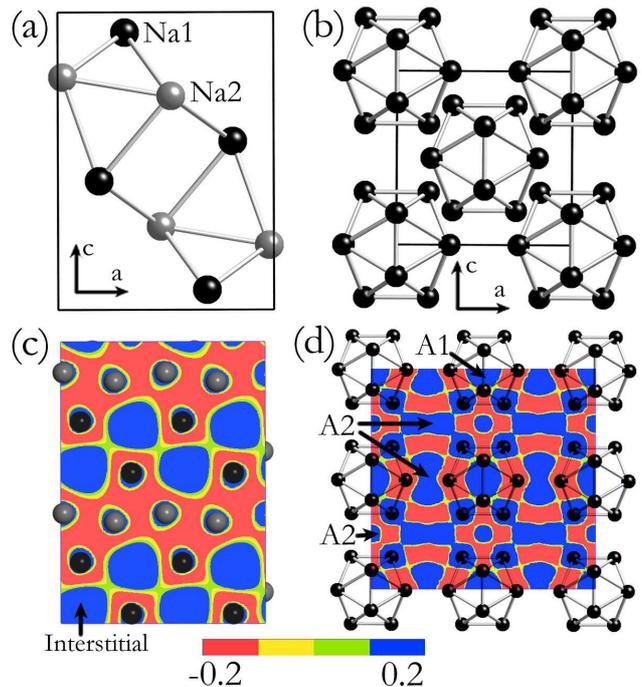}
  \caption{\label{fig:1} (color online) Crystal structures of oP8 (a) and cI24 (b) and their electron density difference maps (c-d) at ultra-high pressures. The lattice parameters of oP8-Na at 2 TPa are \emph{a} = 2.88 ${\AA}$, \emph{b} = 2.12 ${\AA}$, and \emph{c} = 4.0 ${\AA}$ with two inequivalent Na1 and Na2 atoms at 4\emph{c} positions of (0.809, 0.75, 0.567) and (0.523, 0.25, 0.719), respectively. For cI24-Na at 16 TPa, \emph{a} = 3.12 ${\AA}$ with Na atoms at 24\emph{g} positions (0, 0.302, 0.183). Electron density differences are plotted within the (010) plane of oP8-Na (c) at 2 TPa and cI24-Na (d) at 16 TPa, respectively. \emph{A}1 and \emph{A}2 in (d) represent electron attractors located at the centers of the Na$_{12}$ icosahedra and octahedral voids formed by the six neighboring Na$_{12}$ icosahedra, respectively.}
  \end{center}
\end{figure}

Our structure searches readily reproduced the known double hexagonal hP4 structure (space group P6$_3$/mmc, 4 atoms/unit cell) above 400 GPa and we found no better structure below 1.5 TPa. Surprisingly, above 2 TPa, we found the reappearance of the orthorhombic oP8 structure (space group Pnma, 8 atoms/unit cell, Fig. \ref{fig:1}a) which is also observed at low pressures of 117-125 GPa\cite{24}. Strikingly, we predict a hitherto unexpected highly symmetric bcc structure at 20 TPa (space group Im-3, 24 atoms/unit cell, denoted cI24 hereafter, see Fig. 1b). The cI24 structure consists of Na$_{12}$ cages, each of which contains 12 Na atoms forming an icosahedron. The shortest intra- and inter-icosahedral Na-Na distances at 16 TPa are calculated to be 1.138 {\AA} and 1.168 {\AA}, respectively, indicating fairly strong core-core overlap (the ionic radius of Na$^{+}$ is 1.02 {\AA}). An analysis of electron density differences (Fig. \ref{fig:1}d) reveals that the cI24 phase is also an electride with two distinct electron attractors located at the centers of the Na$_{12}$ icosahedra (\emph{A}1) and the octahedral voids (\emph{A}2). However, unlike the nearly spherical isosurface of the interstitial electrons in the hP4 and oP8 structures, dumbbell-like electron localization is found in the octahedral voids.

\begin{figure}
  \begin{center}
  \includegraphics[width=1.0\columnwidth]{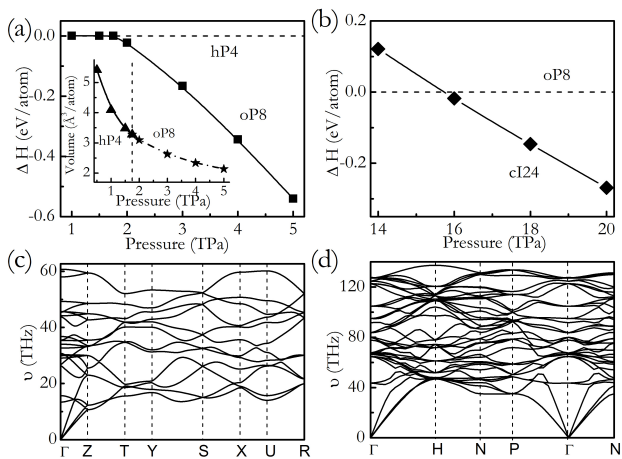}
  \caption{\label{fig:2} (color online) Enthalpy differences (relative to hP4 (a) and oP8 (b)) of Na as a function of pressure. The inset in (a) shows the pressure dependence of the volumes of the hP4 and oP8 phases. The predicted sequence of phase transitions at zero temperature in the pressure range of 0 - 20 TPa is provided in supplementary Figure S1. The volume appears to change continuously across the phase transition. (c) and (d) show the calculated phonon dispersion relations of oP8-Na at 2 TPa and cI24-Na at 16 TPa, respectively. }
  \end{center}
\end{figure}

The enthalpies of the oP8 (relative to hP4) and cI24 (relative to oP8) phases are plotted as functions of pressure in Figs. \ref{fig:2} (a) and (b), respectively. The inset to Fig. \ref{fig:2} (a) shows that hP4-Na transforms to the oP8 structure at ~1.75 TPa with a continuous volume change, indicating a second-order phase transition. The oP8 phase remains stable up to 15.5 TPa, above which the cI24 phase becomes the most stable. Phonon calculations for oP8 and cI24-Na in their corresponding stable pressures ranges show no imaginary frequencies, demonstrating the dynamical stability of the structures (Fig. \ref{fig:2}c, d). The maximum frequency in the TPa pressure regime reaches ~137 THz at 16 TPa, which is very high and implies a large nuclear zero-point (ZP) energy. The estimated ZP energies within the quasi-harmonic approximation are indeed extremely large: 0.577 and 0.585 eV/atom for cI24 and oP8 phases at 20 TPa, respectively. However, the difference in ZP energies of only 8 meV/atom is too small to modify the phase diagram of Na at TPa pressures. Furthermore, we have calculated the Gibbs free energies of the cI24 and oP8 phases at 20 TPa within the quasi-harmonic approximation at finite temperatures to account for vibrational contributions. The resultant difference in free energy between cI24 and oP8 at 1000 K is only 3 meV/atom larger than that at 0 K (Fig. \ref{fig:2}b), which gives a negligible change in the oP8 - cI24 transition pressure.

\begin{figure}
  \begin{center}
  \includegraphics[width=1.0\columnwidth]{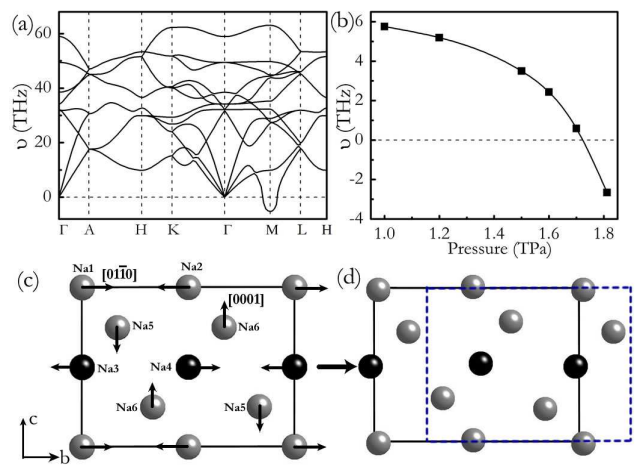}
  \caption{\label{fig:3} (color online) Phonon dispersion curves of hP4-Na at 2 TPa (a) and the phonon frequencies of the softened TA phonon mode at \emph{M} (0, 0.5, 0) as a function of pressure (b). (c) shows the atomic vibrations of the soft TA mode at the \emph{M} point viewed along the \emph{a}-axis in a 1$\times$2$\times$1 supercell of the hP4 structure. Here, the Na1 (Na2) and Na4 (Na3) atoms vibrate parallel (antiparallel) to the [01\ $\bar{1}$0] direction, while the Na5 (Na6) atoms antiparallel (parallel) to the [0001] direction. The gray and black balls indicate atoms in different planes. Arrows indicate vibrational direction of different Na atoms as labeled and are plotted from the eigenvector of the TA soft mode at the M point. The blue dashed cell in (d) indicates the oP8 structure resulting from the distortion.}
  \end{center}
\end{figure}

We calculated the phonon dispersion curves of hP4-Na at various pressures in order to understand the origin of the re-emergence of the oP8 phase and the second-order nature of the hP4 - oP8 transition. We found that a transverse acoustic (TA) phonon mode at the zone boundary \emph{M} (0 0.5 0) point softens with increasing pressure (Fig. 3a) and its frequency goes to zero at ~1.72 TPa (Fig. 3b). A frozen-phonon calculation was performed by distorting the hP4 structure according to the atomic vibrations for the TA mode at the \emph{M} point as depicted in Fig. \ref{fig:3}(c), and structural optimizations were subsequently carried out. The resultant energetically stable structure is indeed the orthorhombic oP8 phase, as indicated by the dashed cell in Fig. \ref{fig:3} (d). It is clear that the hP4 - oP8 transition is driven by the soft TA mode. This result is consistent with the second-order hP4 - oP8 transition characterized by a continuous volume change (inset in Fig. \ref{fig:2}a). Soft-phonon driven hP4 - oP8 transformations have also been reported in MnAs, CrTe, MnTe, MgTe, CrSb and CaH$_2$\cite{45,46,47}.

\begin{figure}
  \begin{center}
  \includegraphics[width=1.0\columnwidth]{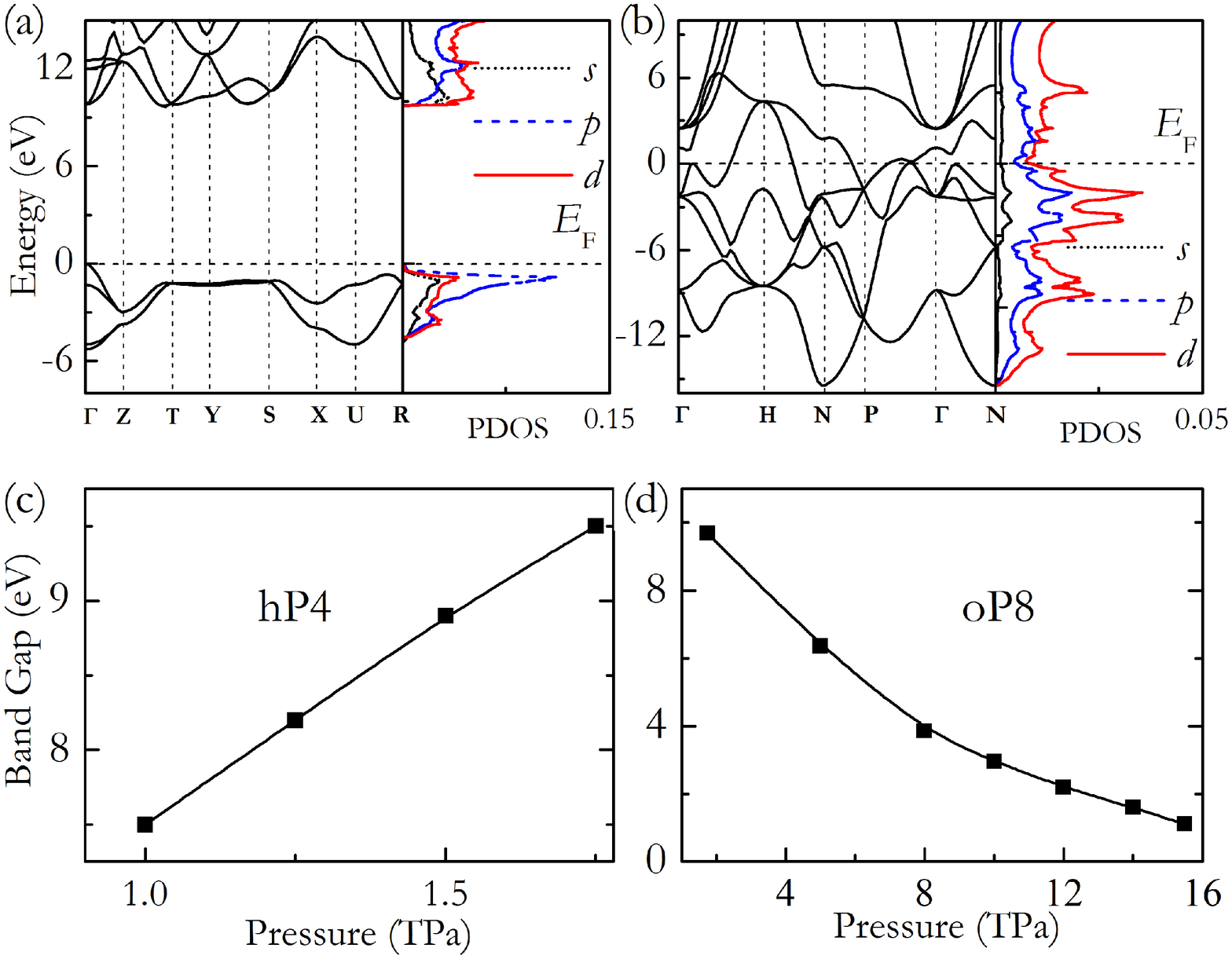}
  \caption{\label{fig:4} (color online) Electronic band structure and projected density of states of oP8-Na at 1.75 (PDOS, in units of eV$^{-1}$ per atom) (a) and cI24-Na at 16 TPa (b). The horizontal dotted lines indicate the Fermi level. (c) and (d) show the variation of band gaps with pressure for hP4 and oP8-Na, respectively, using the HSE06 functional.}
  \end{center}
\end{figure}

The electronic band structures and band gaps of hP4, oP8, and cI24-Na were calculated at selected pressures using the HSE06 functional\cite{48,49}, which is known to give better band gaps than the standard PBE functional. The band gap of hP4-Na increases with pressure and reaches ~9.5 eV at the transition to the oP8 phase at 1.75 TPa. At the transition, oP8-Na is also a wide band gap insulator with an even larger gap of ~9.7 eV (Fig. \ref{fig:4}a). The gap of oP8-Na decreases with pressure down to 1.6 eV at 15.5 TPa. The oP8 structure is compressed by a factor of three from 125 GPa (volume = 77.9 {\AA}$^3$ per unit cell) to 1.75 TPa (volume = 25.9 {\AA}$^3$ per unit cell). As a result of this strong compression, the oP8 structure at TPa pressures forms an electride with valence electrons trapped at the interstitial sites (Fig. \ref{fig:1}c). The electronic structure is fundamentally different from that of the nearly-free-electron behavior of oP8 at a much lower pressure of 125 GPa (supplementary Figure S2)\cite{50}. As a result, the oP8 phase is insulating at TPa pressures, but is metallic at pressure below 125 GPa\cite{24,50}.

After the transition to the cI24 phase, the localized electrons in the octahedral voids are partially squeezed out due to the decreasing interstitial space, forming a series of conducting channels between \emph{A}2 attractors and Na cations, as shown in Fig. \ref{fig:1} (d). Notably, the band structure of cI24-Na at 16 TPa shows highly dispersive bands crossing the Fermi level (Fig. \ref{fig:4}b), and the return to metallicity.

The enthalpy \emph{H} = \emph{U}+\emph{PV} governs the phase stability at 0 \emph{K}.  Under sufficiently strong compression the hard core repulsion between atoms becomes very important and it favors close-packed structures such as (bcc, fcc or hcp)\cite{51} which have small volumes \emph{V}. However, the cI24 structure which Na adopts at TPa pressure is far from simple, and is constructed from Na$_{12}$ icosahedral cages, further challenging the established picture of simple packing at high densities. Indeed, our optimized hypothetical bcc structure with Na atoms occupying the lattice sites at 20 TPa gives a volume of 1.175 {\AA}$^3$/atom, which has a lower density than cI24-Na (1.148 {\AA}$^3$/atom) whose bcc lattice sites are occupied by Na$_{12}$ icosahedral cages. Integration of the electron density at 16 TPa shows that the charge on each Na atom is depleted by about $\sim$0.35 electrons, while the cage-center \emph{A}1 and interstitial \emph{A}2 sites (Fig. \ref{fig:1}d) have excess electronic charges of about $\sim$1.2 and $\sim$1.0 electrons, respectively. As a whole, 9 of the 12 valence electrons remain within each Na$_{12}$ cage (including the localized central electrons), while 3 electrons localize in the octahedral voids. The 9-electron configuration might allow an Na$_{12}$ cage to be termed a super-alkali atom. Electronically, a Na$_{12}$ cage resembles a Potassium atom with an electronic configuration of 3\emph{s}$^2$ 3\emph{p}$^6$ 4\emph{s}$^1$.

Cage-like structures are seldom seen at high pressures since they do not usually favor dense packing. Known cage structures are typically found in elements in which it is easy to form covalent bonds. Examples occur in various allotropes of B ($\alpha$-B$_{12}$\cite{52} and $\gamma$-B$_{28}$\cite{53}) and C (C$_{60}$ and C$_{70}$)\cite{54}, and Si/Ge cages in clathrates (e.g., in Ba$_8$Si$_{46}$\cite{55}). In recent work we also found an N$_{10}$ cage structure in polymeric nitrogen\cite{56} and an H$_{24}$ cage in hydrogen-rich CaH$_6$\cite{35}. Here, the formation of Na$_{12}$ icosahedra is rather surprising since Na is unable to stabilize covalent bonds. The Na$_{12}$ cage is geometrically similar to a B$_{12}$ cage, which is the fundamental building block of various boron allotropes\cite{52,53}. B is electron deficient and forms multi-center covalent bonds. Three-center covalent bonds within the icosahedra and two/three-center covalent bonds between the icosahedra in various structures of elemental B satisfy the octet rule and generate insulating states. Here, there is no covalent bonding within the Na$_{12}$ cages and it is actually a well-packed ionic structure. The electrons at the center of the cage contribute to the stability of the Na$_{12}$ cages, and play a similar role to that of Ba atoms in the Ba$_8$Si$_{46}$ clathrate structure\cite{55}.

We find that cI24-Na is structurally similar to the bcc form of the Al$_{12}$W alloy\cite{57}, in which Al$_{12}$ also forms an icosahedral cage with a W atom localized at its center, and the W-encaged Al$_{12}$ icosahedrons occupy bcc lattice sites. One Al$_{12}$W unit resembles a Na$_{12}$ icosahedron, and the encaged electron plays the role of the pseudo-ions in W. In contrast to the single dumbbell-like electron attractor found in the octahedral voids of cI24-Na, the voids in Al$_{12}$W contain four distinct dumbbell-like electron attractors (supplementary Fig. S5)\cite{50}.

Na is insulating over the large pressure range of 0.2-15.5 TPa. The pressure interval of 15.3 TPa (98.7\% of the pressure range 0 - 15.5 TPa) is comparable to the 17 TPa (33.3\% of the whole pressure range) pressure interval predicted for the insulating phase of Ni\cite{Ni}. The pressure for re-entrant metallicity in Na (15.5 TPa) is much higher than in Li (120 GPa), which may be associated with the entirely different mechanism for insulating behavior. In the hP4 and oP8 structures of Na, the valence electrons localize strongly within the lattice interstices and maintain a nearly spherical shape (Fig. \ref{fig:1}c) up to 15.5 TPa. However, semiconducting Li has a complex oC40 structure with interstitial regions of different shapes in which electrons are localized at a rather low pressure of 80 GPa. These localized electrons are connected to each other and form conducting channels at 120 GPa\cite{50}.

Y. L., Y. W., and Y. M. acknowledge funding from China 973 Program under Grant No. 2011CB808200, and the National Natural Science Foundation of China under Grant Nos. 11274136, 11025418, and 91022029. Y. L. also acknowledges funding from the National Natural Science Foundation of China under Grant Nos. 11204111, 11404148,the Natural Science Foundation of Jiangsu province under Grant No. BK20130223, and the PAPD of Jiangsu Higher Education Institutions. C. J. P. and R. J. N. acknowledge funding from the Engineering and Physical Sciences Research Council of the U.K.


\begin{thebibliography}{99}

\bibitem{1}

V. E. Fortov, V. V. Yakushev, K. L. Kagan, I. V. Lomonosov, V. I. Postnov, and T. I. Yakusheva, JETP Lett. \textbf{70}, 628 (1999).

\bibitem{2}
J. Wittig, Phys. Rev. Lett. \textbf{24}, 812 (1970).

\bibitem{3}
V. V. Struzhkin, M. I. Eremets, W. Gan, H. K. Mao, and R. J. Hemley, Science \textbf{298}, 1213 (2002).

\bibitem{4}
C.-S. Zha and R. Boehler, Phys. Rev. B \textbf{31}, 3199 (1985).

\bibitem{5}
E. Gregoryanz, O. Degtyareva, M. Somayazulu, R. J. Hemley, and H.-k. Mao, Phys. Rev. Lett. \textbf{94}, 185502 (2005).

\bibitem{6}
T. Matsuoka and K. Shimizu, Nature \textbf{458}, 186 (2009).

\bibitem{7}
Y. Ma, M. Eremets, A. R. Oganov, Y. Xie, I. Trojan, S. Medvedev, A. O. Lyakhov, M. Valle, and V. Prakapenka,
Nature \textbf{458}, 182 (2009).

\bibitem{8}
J. B. Neaton and N. W. Ashcroft, Nature \textbf{400}, 141 (1999).

\bibitem{9}
J. B. Neaton and N. W. Ashcroft, Phys. Rev. Lett. \textbf{86}, 2830 (2001).

\bibitem{10}
T. Matsuoka, M. Sakata, Y. Nakamoto, K. Takahama, K. Ichimaru, K. Mukai, K. Ohta, N. Hirao, Y. Ohishi, and K. Shimizu, Phys. Rev. B \textbf{89}, 144103 (2014).

\bibitem{11}
M. Marqu\'{e}s, M. I. McMahon, E. Gregoryanz, M. Hanfland, C. L. Guillaume, C. J. Pickard, G. J. Ackland, and R. J. Nelmes, Phys. Rev. Lett. \textbf{106}, 095502 (2011).

\bibitem{12}
J. Lv, Y. Wang, L. Zhu, and Y. Ma, Phys. Rev. Lett. \textbf{106}, 015503 (2011).

\bibitem{13}
Y. Yao, J. S. Tse, and D. D. Klug, Phys. Rev. Lett. \textbf{102}, 115503 (2009).

\bibitem{14}
C. J. Pickard and R. J. Needs, Phys. Rev. Lett. \textbf{102}, 146401 (2009).

\bibitem{15}
M. Marqu\'{e}s, M. Santoro, C. L. Guillaume, F. A. Gorelli, J. Contreras-Garc\'{\i}a, R. T. Howie, A. F. Goncharov,
and E. Gregoryanz, Phys. Rev. B \textbf{83}, 184106 (2011).

\bibitem{16}
C. L. Guillaume, E. Gregoryanz, O. Degtyareva, M. I. McMahon, M. Hanfland, S. Evans, M. Guthrie, S. V. Sinogeikin, and H. K. Mao, Nat. Phys. \textbf{7}, 211 (2011).

\bibitem{17}
M.-S. Miao and R. Hoffmann, Acc. Chem. Res. \textbf{47}, 1311 (2014).

\bibitem{18}
C. J. Pickard and R. J. Needs, Nat. Mater. \textbf{9}, 624 (2010).

\bibitem{19}
P. Li, G. Gao, Y. Wang, and Y. Ma, J. Phys. Chem. C \textbf{114}, 21745 (2010).

\bibitem{20}
A. R. Oganov, Y. Ma, Y. Xu, I. Errea, A. Bergara, and A. O. Lyakhov, Proc. Nat. Acad. Sci. U.S.A. \textbf{107}, 7646 (2010).

\bibitem{21}
C. J. Pickard and R. J. Needs, Phys. Rev. Lett. \textbf{107}, 087201 (2011).

\bibitem{Mg}
J. A. Moriarty and A. K. McMahan, Phys. Rev. Lett. \textbf{48}, 809 (1982).

\bibitem{Ni}
A. K. McMahan and R. C. Albers, Phys. Rev. Lett. \textbf{49}, 1198 (1982).

\bibitem{22}
E. Wigner and H. Huntington, J. Chem. Phys. \textbf{3}, 764 (1935).

\bibitem{23}
M. Hanfland, I. Loa, and K. Syassen, Phys. Rev. B \textbf{65}, 184109 (2002).

\bibitem{24}
E. Gregoryanz, L. F. Lundegaard, M. I. McMahon, C. Guillaume, R. J. Nelmes, and M. Mezouar, Science \textbf{320}, 1054 (2008).

\bibitem{25}
M. I. McMahon et al., Proc. Nat. Acad. Sci. U.S.A. \textbf{104}, 17297 (2007).

\bibitem{26}
M. Hanfland, K. Syassen, L. Loa, N. E. Christensen, and D. L. Novikov, Poster at 2002 High Pressure Gordon Conference  (2002).

\bibitem{27}
Y. Wang, J. Lv, L. Zhu, and Y. Ma, Phys. Rev. B \textbf{82}, 094116 (2010).

\bibitem{28}
Y. Wang, J. Lv, L. Zhu, and Y. Ma, Comput. Phys. Commun. \textbf{183}, 2063 (2012).

\bibitem{29}
C. J. Pickard and R. J. Needs, Phys. Rev. Lett. \textbf{97}, 045504 (2006).

\bibitem{30}
C. J. Pickard and R. J. Needs, J. Phys.: Condens. Matter \textbf{23}, 053201 (2011).

\bibitem{31}
Q. Li, D. Zhou, W. Zheng, Y. Ma, and C. Chen, Phys. Rev. Lett. \textbf{110}, 136403 (2013).

\bibitem{32}
L. Zhu, Z. Wang, Y. Wang, G. Zou, H. Mao, and Y. Ma, Proc. Nat. Acad. Sci. U.S.A. \textbf{109}, 751 (2011).

\bibitem{33}
L. Zhu, H. Wang, Y. Wang, J. Lv, Y. Ma, Q. Cui, Y. Ma, and G. Zou, Phys. Rev. Lett. \textbf{106}, 145501 (2011).

\bibitem{34}
Y. Wang, H. Liu, J. Lv, L. Zhu, H. Wang, and Y. Ma, Nat. Commu. \textbf{2}, 563 (2011).

\bibitem{35}
H. Wang, S. T. John, K. Tanaka, T. Iitaka, and Y. Ma, Proc. Nat. Acad. Sci. U.S.A. \textbf{109}, 6463 (2012).

\bibitem{36}
C. J. Pickard and R. J. Needs, J. Chem. Phys. \textbf{127}, 244503 (2007).

\bibitem{37}
C. J. Pickard and R. J. Needs, Nat. Phys. \textbf{3}, 473 (2007).

\bibitem{38}
J. M. McMahon and D. M. Ceperley, Phys. Rev. Lett. \textbf{106}, 165302 (2011).

\bibitem{39}
C. J. Pickard, M. Martinez-Canales, and R. J. Needs, Phys. Rev. B \textbf{85}, 214114 (2012).

\bibitem{40}
J. Sun, M. Martinez-Canales, D. D. Klug, C. J. Pickard, and R. J. Needs, Phys. Rev. Lett. \textbf{108}, 045503 (2012).

\bibitem{41}
R. Smith, J. Eggert, R. Jeanloz, T. Duffy, D. Braun, J. Patterson, R. Rudd, J. Biener, A. Lazicki, and A. Hamza, Nature \textbf{511}, 330 (2014).

\bibitem{42}
S. Baroni, P. Giannozzi, and A. Testa, Phys. Rev. Lett. \textbf{58}, 1861 (1987).

\bibitem{43}
S. J. Clark, M. D. Segall, C. J. Pickard, P. J. Hasnip, M. I. Probert, K. Refson, and M. C. Payne, Z. Kristallogr. \textbf{220}, 567 (2005).

\bibitem{44}
J. P. Perdew, K. Burke, and M. Ernzerhof, Phys. Rev. Lett. \textbf{77}, 3865 (1996).

\bibitem{45}
Y. W. Li, B. Li, T. Cui, Y. Li, L. J. Zhang, Y. M. Ma, and G. T. Zou, J. Phys.: Condens. Matter \textbf{20}, 045211 (2008).

\bibitem{46}
Y. Li, Y. Ma, T. Cui, Y. Yan, and G. Zou, Appl. Phys. Lett. \textbf{92}, 101907 (2008).

\bibitem{47}
Y. Li, Y. W. Li, Y. Ma, T. Cui, and G. Zou, Phys. Rev. B \textbf{81}, 052101 (2010).

\bibitem{48}
J. Heyd, G. E. Scuseria, and M. Ernzerhof, J. Chem. Phys. \textbf{118}, 8207 (2003).

\bibitem{49}
A. V. Krukau, O. A. Vydrov, A. F. Izmaylov, and G. E. Scuseria, J. Chem. Phys. \textbf{125}, 224106 (2006).

\bibitem{50}
See the Supplemental Material at http://link.aps.org/ 414 supplemental/10.1103/PhysRevLett.000.000000 for phase sequence, band structures, density of states, and electron density difference.

\bibitem{51}
B. Rousseau, Y. Xie, Y. Ma, and A. Bergara, Eur. Phys. J. B \textbf{81}, 1 (2011).

\bibitem{52}
A. Masago, K. Shirai, and H. Katayama-Yoshida, Phys. Rev. B \textbf{73}, 104102 (2006).

\bibitem{53}
A. R. Oganov, J. Chen, C. Gatti, Y. Ma, C. W. Glass, Z. Liu, T. Yu, O. O. Kurakevych, and V. L. Solozhenko, Nature \textbf{457}, 863 (2009).

\bibitem{54}
W. Kr\"{a}tschmer, L. D. Lamb, K. Fostiropoulos, and D. R. Huffman, Nature \textbf{347}, 354 (1990).

\bibitem{55}
S. Yamanaka, E. Enishi, H. Fukuoka, and M. Yasukawa, Inorg. Chem. \textbf{39}, 56 (2000).

\bibitem{56}
X. Wang, Y. Wang, M. Miao, X. Zhong, J. Lv, T. Cui, J. Li, L. Chen, C. J. Pickard, and Y. Ma, Phys. Rev. Lett. \textbf{109}, 175502 (2012).

\bibitem{57}
W. B. Pearson, Crystal chemistry and physics of metals and alloys (Wiley, New York, 1972).

\end{thebibliography}
\end{document}